\documentclass[showpacs,amsmath,amssymb,aps,floatfix]{revtex4}

\usepackage{graphicx,color}
\usepackage{dcolumn}
\usepackage{bm}

\begin{document}

\preprint{paperPABIO10 - GO net}

\title{Gene-based and semantic structure of the Gene Ontology as a complex network}
\author{S. Miccich\`e}
\affiliation{Universit\`a degli Studi di Palermo, Dipartimento di Fisica\\
                   Viale delle Scienze, Ed. 18, I-90128 Palermo, Italy}

\begin{abstract} 

\noindent The last decade has seen the advent and consolidation of ontology based tools for the identification and biological interpretation of classes of genes, such as the Gene Ontology. The Gene Ontology is constantly evolving over time. The information accumulated time-by-time and included in the GO is encoded in the definition of terms and in the setting up of semantic relations amongst terms. This approach might be usefully complemented by a bottom-up approach based on the knowledge of relationships amongst genes. To this end, ee investigate the Gene Ontology from a complex network perspective. We consider the semantic network of terms naturally associated with the semantic relationships provided by the Gene Ontology consortium and a  gene-based weighted network in which the nodes are the terms and a link between any two terms is weighted according to the number of genes that are annotated in both terms. One aim of the present paper is to understand whether the semantic and the gene-based network share the same structural properties or not. Indeed, we will show that the main differences between the two networks are in the number of links and not in the relative importance of the terms within the network. We then consider network communities. We show that in some cases GO terms that appear to be distinct from a semantic point of view are instead connected when considering their gene content.  The identification of communities in the SVNs network can therefore be the basis of a simple protocol aiming at fully exploiting the possible relationships amongst terms, thus improving the semantic structure of GO. However, this is also important from a biomedical point of view, as it might reveal how genes over-expressed in a certain term also affect other biological processes, molecular functions and cellular components not directly linked by the GO semantics. As a by-product, we present a simple methodology that allows to have a first glance insight about the biological characterization of groups of GO terms.

\end{abstract}


\date{\today}

\maketitle

\section{Introduction} \label{intro}

The last decade has seen the advent and consolidation of ontology based tools for the identification and biological interpretation of classes of genes, such as the Gene Ontology (GO) \cite{GOcons}. Typically, ontologies allow for associating a gene to its biological functions and they also provide the information about the other genes which cooperate in performing such functions. As such, ontologies are a useful tool for exploiting the existence of sets of genes involved in a certain pathology. These instruments are alternative to the usual clustering methodologies, which are for example used in the analysis of gene expression profiles obtained from microarray experiments. One main difference is that the classes of genes obtained in an ontological analysis have a clear biological interpretation. 

The GO is constantly evolving \cite{leonelli,alterowitz,diehl} over time. The information accumulated time-by-time and included in the GO is encoded in the definition of terms and in the setting up of relations amongst terms. Thus a key point for the evolution and maintenance of GO is to set up protocols that are able to capture the relations amongst genes so that they can be profitably transferred at the level of terms. In fact, the GO is a controlled vocabulary where not only each terms is explained in some detail, but it is also given a set of relations between the terms. This semantic structure is mainly based on biological evidences relative to the functions described by the terms, i.e. on the knowledge of existing relations amongst biological functions.

This is a top-down approach that we think might be usefully complemented by a bottom-up approach based on the direct knowledge of relationships amongst genes. Given a set of terms, one can set up a link between any two terms if they have a gene in common. Therefore, in parallel with the semantic network of terms naturally associated with the semantic relationships provided by GO, it is possible to generate a  gene-based weighted network in which the nodes are the terms and a link between any two terms is weighted according to the number of genes that are annotated in both terms \cite{paperbio8}. Moreover, it is possible to consider the bipartite system of GO terms and genes and investigate its properties by constructing and analyzing the projected network on one of the two sets. Here we will be interested on the projected network of terms. Recently a methodology has been proposed that  identifies preferential links in the projected network \cite{SVN}, i.e.  links whose presence in the projected network cannot be explained in terms of a random co-occurrence of neighbors in the bipartite system. The resulting network where the existence of each link has been statistically tested against a null hypothesis of randomness is called statistically validated network (SVN). One aim of the present paper is to understand whether the semantic and the gene-based network share the same structural properties or not. Indeed, we will show that the scatter-plots of the betweenness {\em{versus}} the degree in both networks are compatible with a parabolic dependence of the betweenness with respect to the degree, thus indicating that the main differences between the two networks are in the number of links and not in the relative importance of the terms within the network.

Another way to compare the information encoded in the semantic GO structure with the one associated to the genes annotated in the terms is to investigate the network communities. In fact, once the gene-based network has been constructed, it can be partitioned in order to look for communities. In this case, communities are sets of GO terms that share a similar profile in terms of their annotated genes. The same can be done for the semantic network, thus obtaining sets of GO terms that share a similar profile in terms of their semantic relationships. Indeed, the idea of searching for communities or clusters of GO terms is not new. However, one usually looks for communities or clusters within the semantic GO network only \cite{resnik,lee,speer,wang,frohlich,othman,Gsesame, david}. Our approach is different. In fact, we use the information on the gene content of any GO term in order to create a statistically validated network of GO terms and then we partition it by using any standard community detection algorithm. As we will show below, it turns out that in some cases GO terms present in the same community of the gene-based network are not joined by any semantic link. This shows that terms that appear to be distinct from a semantic point of view are instead connected when considering their gene content.  The identification of communities in the gene-based network can therefore be the basis of a simple protocol able to fully exploit the possible relationships amongst terms, thus improving the semantic structure of GO.  

As a by-product, we present a simple methodology that allows to have a first glance insight about the biological meaning of groups of GO terms. We have put on a statistical basis what any researcher first does when he obtains a list of GO terms that are somehow relevant in the analysis he is performing. The first thing to do is to read the definitions of the GO terms trying to figure out a possible ``story'' for the reason why the obtained terms are clustered together. We have devised a procedure that helps in this direction by providing the most relevant ``words'' of the  ``story''.

The paper is organized as follows: in section \ref{GOreleases} we illustrate the data considered in our investigation while in section \ref{SVNs} we will briefly review the methodologies that allows the generation of statistically validated networks and the cluster characterization. In section \ref{results} we will study the semantic and gene-based network and show GO terms that belong to the same gene-based network community and are not joined by any semantic link. Our conclusions are drawn in section \ref{concl}.

\section{Data and methodology} \label{section2}

\subsection{The Gene Ontology database} \label{GOreleases}

We consider only the human part of the Gene Ontology. To this end we downloaded the ${\tt gene\_association.goa\_human.gz}$ \cite{go} file, release 1.224 with GOC validation date 20/02/2012. This is the file that accounts for the association between terms and genes. Together with that we have also downloaded the ${\tt gene\_ontology\_edit.obo}$ \cite{gobo} file, release 1.1.2667 with release date: 02/03/2012 09:20. This is the files that contains a description of the terms and the semantic links amongst them. Based on the semantic links we associated to a term all gene directly annotated in its children terms. As a result our system involves 12564 terms and 18992 genes.

\subsection{Statistically validated networks} \label{SVNs}

Many complex systems present an intrinsic bipartite nature and are often described and modeled in terms of networks. Bipartite networks are composed by two disjoint sets of nodes, say set ${\cal A}$ and set ${\cal B}$, such that every link connects a node in the first set with a node of the second set. The bipartite network is often transformed by one-mode projecting, i.e. one creates a network of nodes belonging to one of the two sets and two nodes are connected when they have at least one common neighboring node of the other set. 

Bipartite networks  are often very heterogeneous in the number of relationships that the elements of one set establish with the elements of the other set. When one constructs a projected network with nodes from only one set, the system heterogeneity makes it very difficult to identify preferential links between the elements. A new methodology to statistically validate each link of the projected network against a null hypothesis taking into account the heterogeneity of the system has been recently introduced \cite{SVN}. 

Let us consider two nodes $A_1$ and $A_2$ both belonging to ${\cal A}$. Let $N_1$ be the number of elements in set $\cal B$ linked to node $A_1$ and $N_2$ the number of elements in set $\cal B$ linked to node $A_2$. The total number elements in set $\cal B$ is $N_B$ and the observed number of elements in set $\cal B$ both linked to $A_1$ and $A_2$ is $N_{12}$. Under the null hypothesis of random co-occurrence, the probability of observing $X$ co-occurrences of links both in $A_1$ and $A_2$ is given by the hypergeometric distribution \cite{Feller}
\begin{equation}
                                H(X|N_B,N_1,N_2)=\frac{{N_2 \choose X} {N_B-N_2 \choose N_1-X}}{{N_B \choose N_1}}.     \label{hyper}
\end{equation}
We can therefore associate a $p$-value to the observed $N_{12}$ as:
\begin{equation}
                                 p(N_{12}) =1- \sum_{X=0}^{N_{12}-1} H(X|N_B,N_1,N_2)
\end{equation}
It is therefore possible to associate a p-value to each link in the projected network. After fixing a threshold one is thus able to select those links whose p-value is below the threshold. These links constitute the statistically validated network. Our working hypothesis is that such links are the most informative links in the original projected network because they are not compatible with a null hypothesis of randomness.

The selection of the appropriate threshold is a key point. In fact, since the null hypothesis is tested for all links of the original projected network, we are in the typical situation when multiple test correction procedures must be applied. There are two possible correction procedure: the Bonferroni correction \cite{Miller1981} and the FDR correction \cite{Benjamini}, that is less restrictive. The Bonferroni correction is based on the consideration that if one tests $N_t$ either dependent or independent hypotheses on a set of data, then a conservative way of maintaining the error rate low is to test each individual hypothesis at a statistical significance level of $p_t/N_t$, where $p_t$ is the chosen statistical threshold ($1 \%$ in the present study). The threshold of the FDR correction linearly increases with the number of tests in which the null hypothesis is rejected. 

When the heterogeneity in one of the two sets is large, the above approach can be modified as follows. Suppose one wants to validate the links in the set A projected network. If the heterogeneity in the elements of set B is high one can construct bipartite subsystems $S_k$ of the original bipartite system $S$ consists of all the $N_B^k$ elements of set B with a given degree k and of all the elements from set A linked to them. By construction, a subsystem $S_k$ is homogeneous with respect to the degree of elements belonging to set B. The methodology sketched above can therefore be applied to each subsystem $S_k$, thus obtaining a collection of statistically validated networks. In this case the p-value threshold must take into account that we are testing the null hypothesis for each subsystem and for each link in it. We then aggregate all validations by generating a statistically validated network where a link between node pairs is established whenever it has been validated in at least one subsystem. Such link can be given a weight equal to the total number of subsystems in which the link itself has been statistically validated.

We refer to the network obtained by using the FDR correction for multiple comparisons as the FDR network. We refer to the network obtained by using the Bonferroni correction for multiple comparisons as the Bonferroni network. By construction, the Bonferroni network is a subgraph of the FDR network, which is a subgraph of the original projected adjacency network. A software to compute the Bonferroni and FDR network is available at the following web-sites: {\tt{http://ocs.unipa.it/validate.html}} and  {\tt{http://ocs.unipa.it/validate-k.html}}.

\subsection{Cluster characterization} \label{cluschar}

Given a network, a first step in the understanding of the represented system is the identification of communities within the network. Communities are sets of nodes that are linked amongst them to a degree which is higher than the one expected on the basis of a null hypothesis of randomness \cite{Fortunato}. This is equivalent to breaking down atomic particles in smaller pieces in order to understand what are their elementary constituents. This step however requires that these elementary constituents be given a name and their features are clearly stated. In other words, when communities are detected, then it is important to characterize them, i.e. to understand what are the main features that explain why nodes are grouped together in a community.

A statistically robust methodology for the community characterization has been given in Ref. \cite{cluschar}. The main idea is to use attributes specific of the nodes involved in the cluster in order to see which attribute is most represented in the community. Suppose to have a community of K nodes. Suppose X out of K nodes are characterized by having a certain attribute A. Suppose that in a network of N elements the attribute A can be associated to M out of N nodes. Then the probability that X is observed by chance is given by the hypergeometric distribution:
\begin{equation}
                                H(X|N,M,K)=\frac{{M \choose X} {N-M \choose K-X}}{{N \choose K}}.     \label{hyperclus}
\end{equation}
For each attribute present in a community and for each community in the network we can therefore have a $p$-value. By considering the multiple hypothesis testing corrections illustrated above we can thus investigate what are the attributes that result to be over-expressed in a community. 

A software to perform the statistical characterization of communities within a network is available at the following web-site: {\tt{http://ocs.unipa.it/characterize.html}}.

\section{Results and Discussion} \label{results}

\subsection{The semantic adjacency networks} \label{comparison}

The semantic structure of the Gene Ontology can be described in terms of an adjacency network where the nodes are the GO terms and the links between terms are provided by the semantic links of the ${\tt gene\_ontology\_edit.obo}$ file. When restricting to the human case, we get a network with $N=12564$ nodes and $L_s=116422$ links. The semantic network is naturally partitioned in three large communities of size $N_1=8118$, $N_2=3336$ and $N_3=1110$. They correspond to the three main {\em{branches}} of GO: {\sl{GO:0008150 (Biological Processes)}}, {\sl{GO:0005575 (Cellular Component)}} and {\sl{GO:0003674 (Molecular Function)}}. The number of links connecting terms inside the three branches are 85447, 18489 and 12486 respectively. 

\begin{figure} 
\begin{center}
             {
              \includegraphics[scale=0.24] {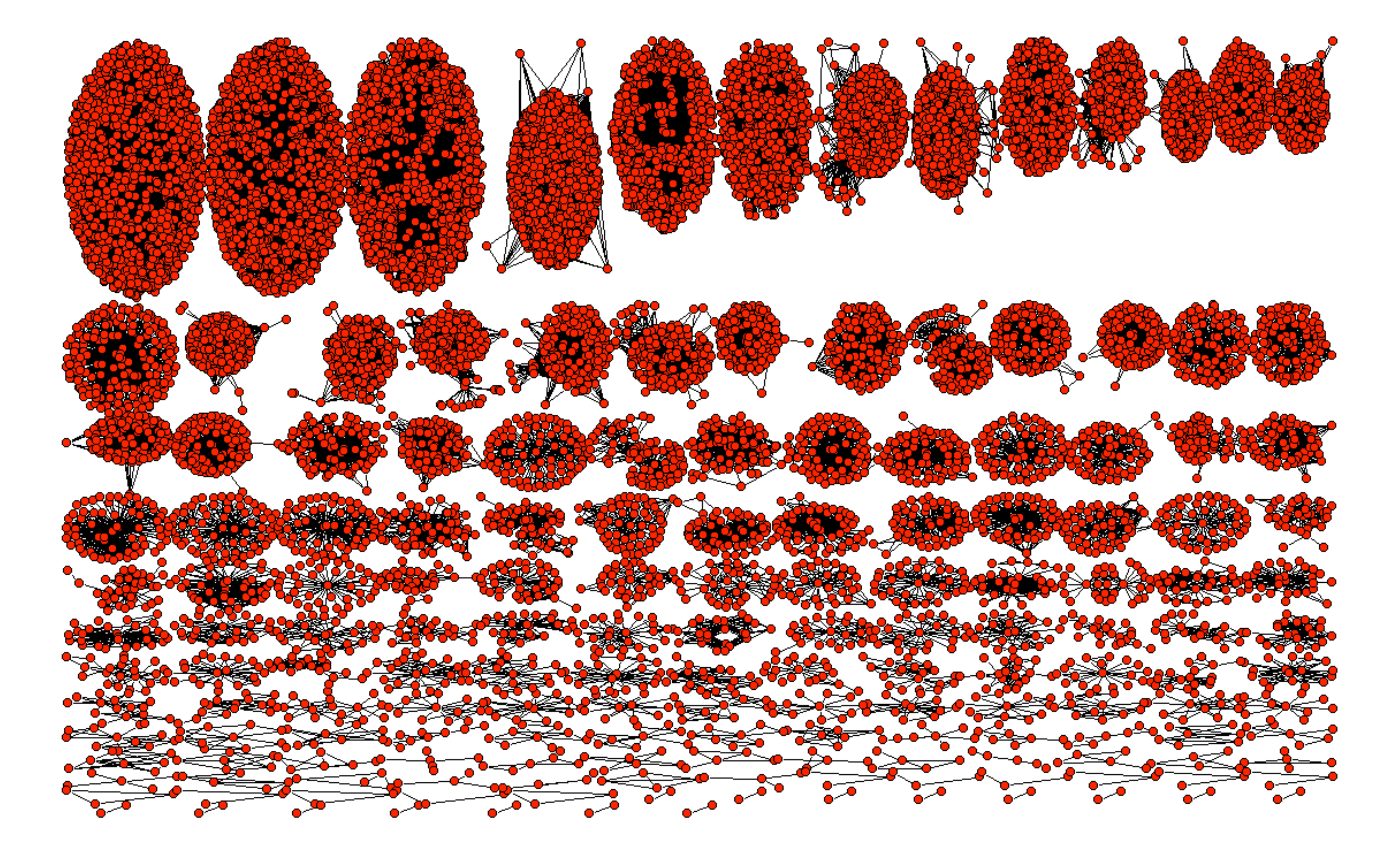} 
              \includegraphics[scale=0.26] {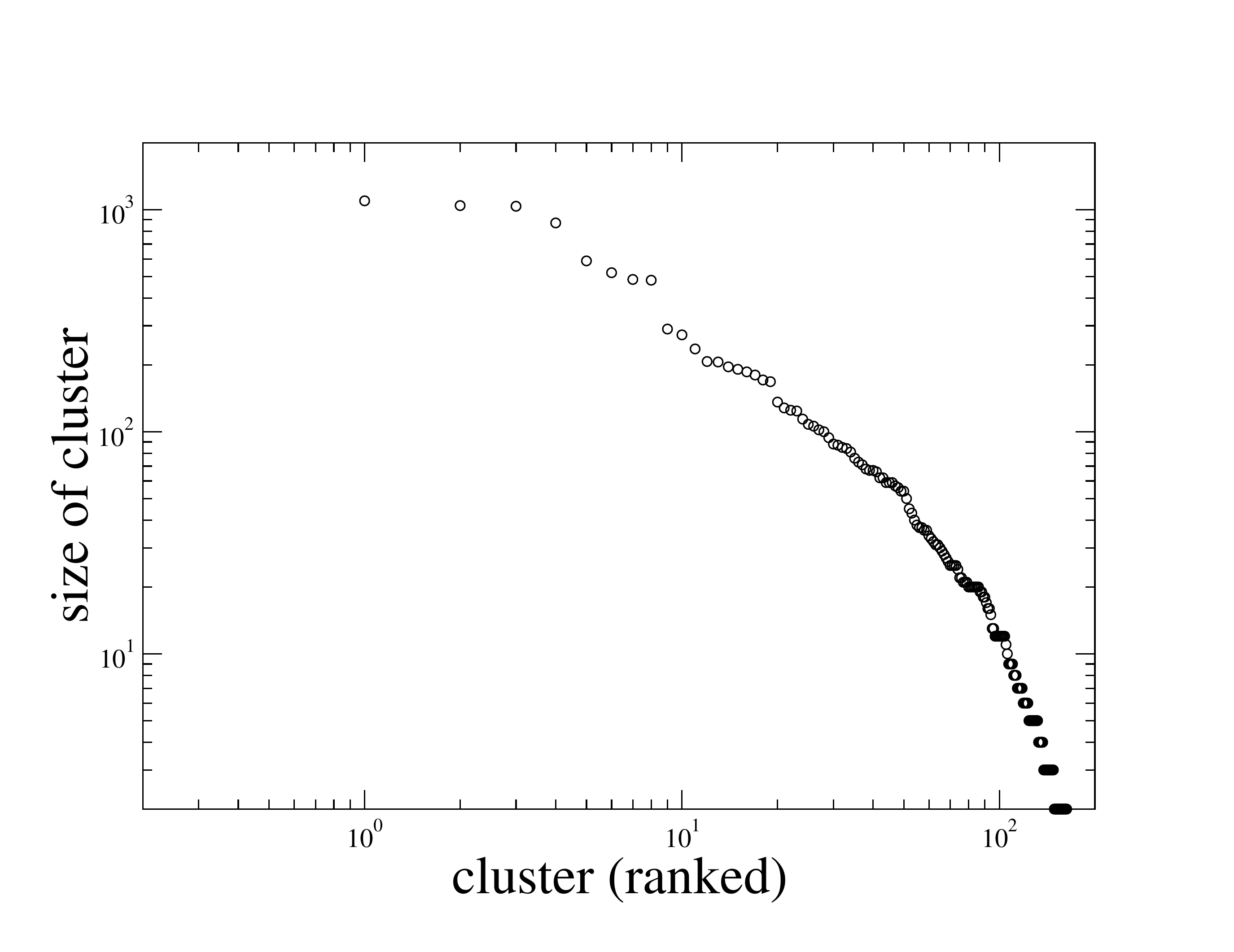} 
             }
             \caption{Partition of the semantic adjacency network by using the Infomap algorithm \cite{Rosvall2008}. The left panel shows the different communities observed. The right panel shows a rank plot of the size of communities.} \label{ADJsemnet}
\end{center}
\end{figure}

The network can be further partitioned by using a community detection algorithm \cite{Fortunato} such as Infomap \cite{Rosvall2008}. We performed 100 runs of the algorithm. In each run the algorithms performed 10 searches. We selected the partition with the lowest {\sl{code length}}. We thus obtained a partition of the adjacency semantic network in $N_p^{(s)}=163$ communities of size ranging from 1096 to 2, as shown in Fig. \ref{ADJsemnet} (panel b). 

The partitioned network is shown in Fig. \ref{ADJsemnet} (panel a). The list with the association between the GO terms and their community is given in Table SM1 of the Supplementary Material. For example the $50^{th}$ largest community, of size 54,  contains terms like {\sl{GO:0001504 (neurotransmitter uptake)}} or {\sl{GO:0007268 (synaptic transmission)}} which are homogeneous from a biological point of view. This points in the direction that the obtained communities are meaningful from a biological point of view. This is confirmed by performing a characterization of the clusters according to the methodology illustrated in Ref. \cite{cluschar}. For each term in  the whole ADJ semantic network we first consider the words that define each term. After eliminating articles and prepositions we have 12855 distinct words. We then construct the 3-words obtained by concatenating together three consecutive words in the terms definition. For example, in the case of the GO term {\sl{GO:0048518 (positive regulation of biological process)}} we get the two 3-words: {\sl{positive\_regulation\_biological}} and {\sl{regulation\_biological\_process}}. We have 34309 distinct 3-words in the whole network. We also delete the 3-words that appear up to two times within the whole network. We therefore have 10254 distinct 3-words. We use these 3-words as attributes to the GO terms and statistically validate the over-expression of them in each of the $N_p^{(s)}=163$ communities obtained by using the Infomap algorithm. All results are given in Table SM2 of the Supplementary Material. For the $50^{th}$ largest community mentioned above the get the following over-represented 3-words: {\sl{regulation\_synapse\_assembly}} and {\sl{positive\_regulation\_synapse}}, thus confirming that the $50^{th}$ largest community groups together GO terms dealing with the regulation of the biological processes active at the level of synaptic transmission. As further examples, in Table \ref{tabsemADJ} we show the results for the first two largest clusters. The first cluster seems to aggregate terms involved in the metabolic processes that break down acids into smaller units and the second cluster groups terms involved in the transport and protein targeting processes.

\begin{table}
\begin{tabular}{|c|c||l|}
\hline
Cluster~ & size & overexpressed 3-words attribute\\
 \hline
$  1$ & $1096$&{\sl{(SSU-rRNA\_5.8S\_rRNA}}\\ 
$  1$ & $1096$&{\sl{5.8S\_rRNA\_LSU-rRNA)}}\\
$  1$ & $1096$&{\sl{sulfate\_proteoglycan\_biosynthetic}}\\ 
$  1$ & $1096$&{\sl{proteoglycan\_biosynthetic\_process}}\\ 
$  1$ & $1096$&{\sl{acid\_catabolic\_process}}\\ 
$  1$ & $1096$&{\sl{mRNA\_catabolic\_process}}\\
$  1$ & $1096$&{\sl{acid\_metabolic\_process}}\\
$  1$ & $1096$&{\sl{compound\_metabolic\_process}}\\
$  1$ & $1096$&{\sl{acid\_biosynthetic\_process}}\\ \hline
$  2$ & $1043$&{\sl{transport\_vesicle\_membrane}}\\ 
$  2$ & $1043$&{\sl{endoplasmic\_reticulum\_membrane}}\\
$  2$ & $1043$&{\sl{side\_plasma\_membrane}}\\
$  2$ & $1043$&{\sl{ubiquitin\_ligase\_complex}}\\ \hline
\end{tabular}  
\caption{Statistical characterization of the communities of the semantic adjacency network. Communities have been obtained by using the Infomap algorithm \cite{Rosvall2008}. The characterization has been done by using the methodology of Ref. \cite{cluschar}. The attributes of each term are the 3-words obtained by concatenating together three consecutive words in the terms definition. Only the results for the first two clusters are shown. The full list of characterizations is given in Table SM1 of the Supplementary Material.} \label{tabsemADJ}
\end{table}

We have therefore devised a simple methodology that allows to have a first glance insight about groups of GO terms. We have put on a statistical basis what any researcher first does when he obtains a list of GO terms that are somehow relevant in the analysis he is performing. The first thing to do is to read the definitions of the GO terms trying to figure out a possible ``story'' for the reason why the obtained terms are together. This procedure described above helps in this direction by providing the most relevant ``words'' of the  ``story''.

\subsection{The gene-based adjacency networks} \label{ADJgene}

\begin{figure} 
\begin{center}
             {
               \includegraphics[scale=0.30] {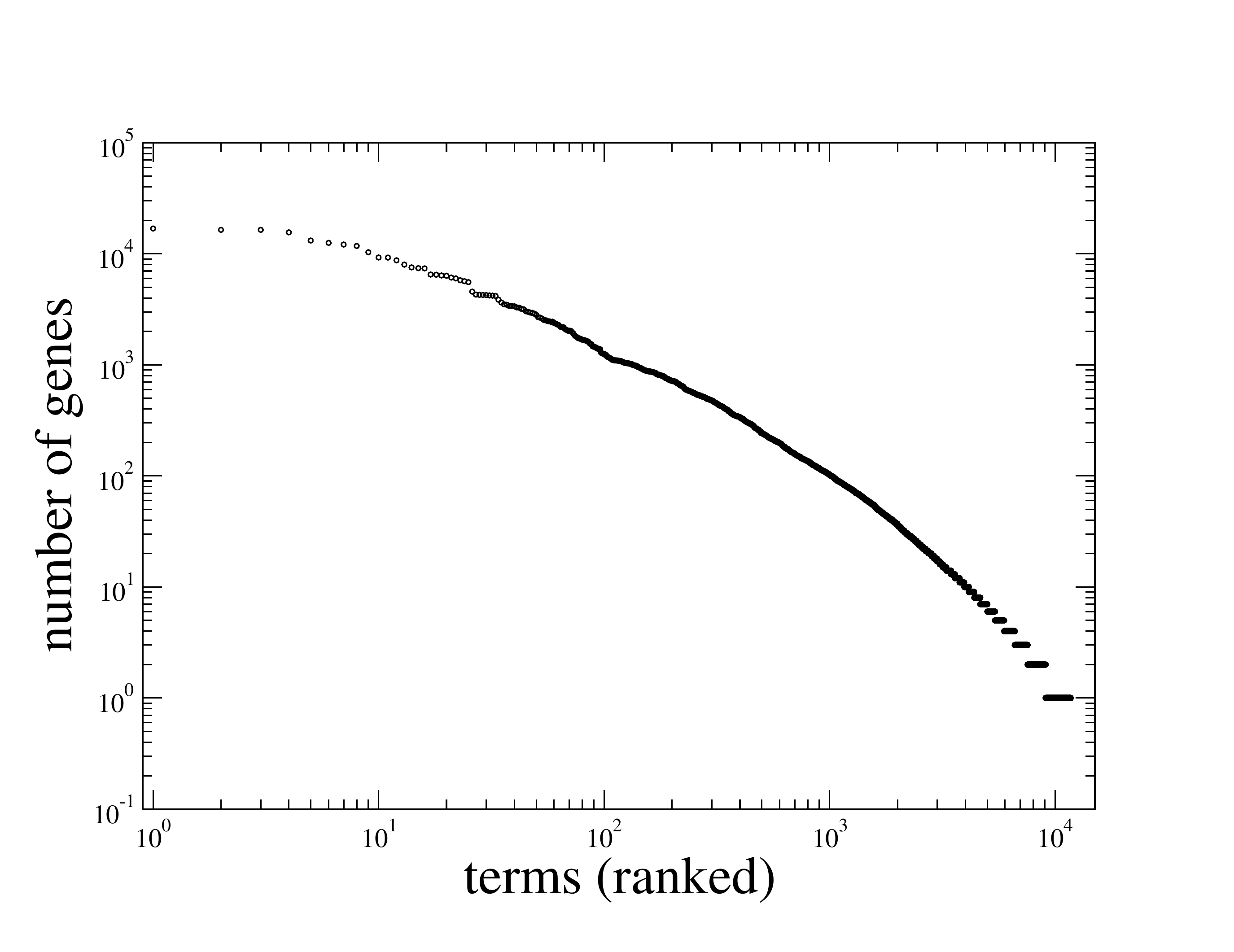} 
               \includegraphics[scale=0.30] {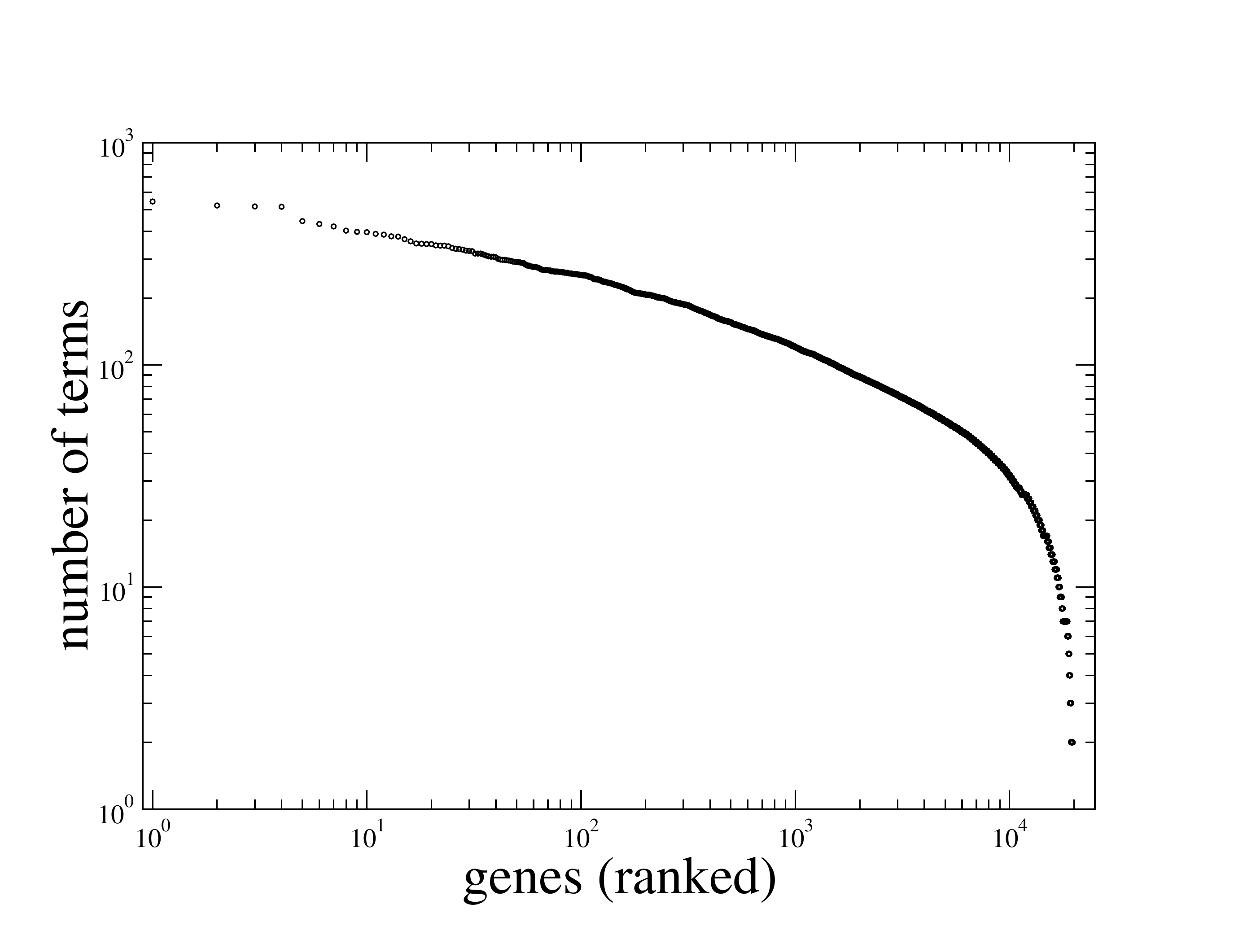}
             }  
              \caption{For the GO terms/genes bipartite system, in the left panel we show the number of genes assigned to any term and in the right panel we show the number of terms a gene belongs to.} \label{GOnetADJ1}
\end{center}
\end{figure}

Let us now consider the GO structure that emerges when considering the gene content of the terms. For each term of the Gene Ontology we consider all genes annotated in it. We assign a gene to a certain term whenever the gene is either directly annotated in it or in any of its children. We thus have a bipartite system of genes and GO terms. In Fig. $(\ref{GOnetADJ1})$ (left panel) we show the number of genes assigned to any term and in Fig. $(\ref{GOnetADJ1})$ (right panel) we show the number of terms a gene belongs to. The heterogeneity is quite large in both respects. There are terms containing over 10000 genes as well as there are gene containing only one gene. Analogously, there are genes present in over 500 GO terms as well as there are genes present in just two terms. 

As mentioned in section \ref{SVNs}, starting from the bipartite system of GO terms and genes it is possible to construct two projected networks: the one of GO terms and the one of genes. The projected network of genes would be a network where nodes are genes and any two genes are connected if they belong to the same GO term. This would provide information about the links between genes based on their membership to biological processes, molecular functions and cellular components. For our purposes we will here consider the projected adjacency network of GO terms, i.e. we can generate the gene-based network where any two terms are connected by a link whenever there exists a least one gene assigned to both terms. Such adjacency network involves $N=12564$ nodes and $L_g=5142743$ links. It is worth noticing that the number of links in this network is much larger than the number $L_s$ of semantic links.  

\begin{figure} 
\begin{center}
               {
                \includegraphics[scale=0.30] {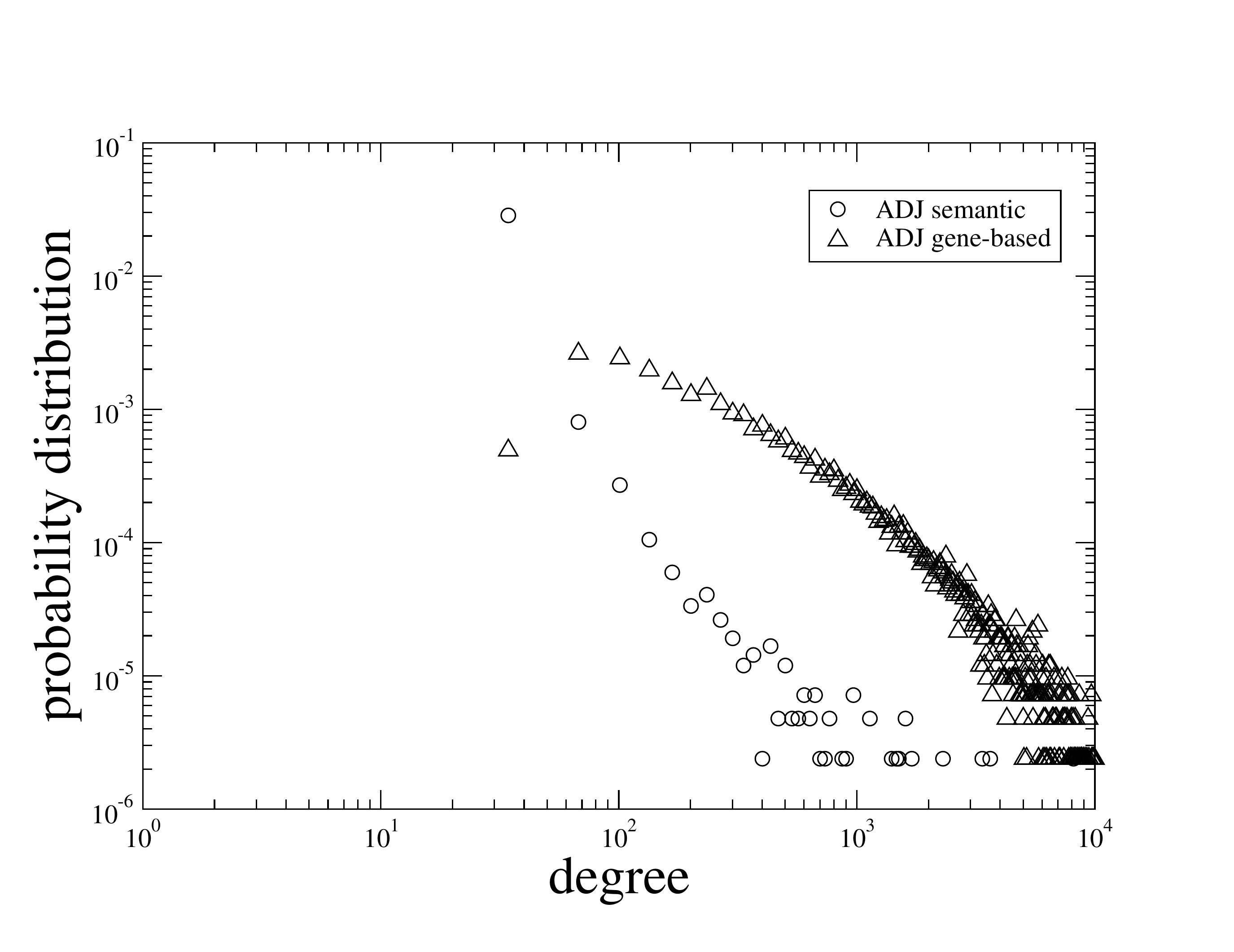} 
                \includegraphics[scale=0.30] {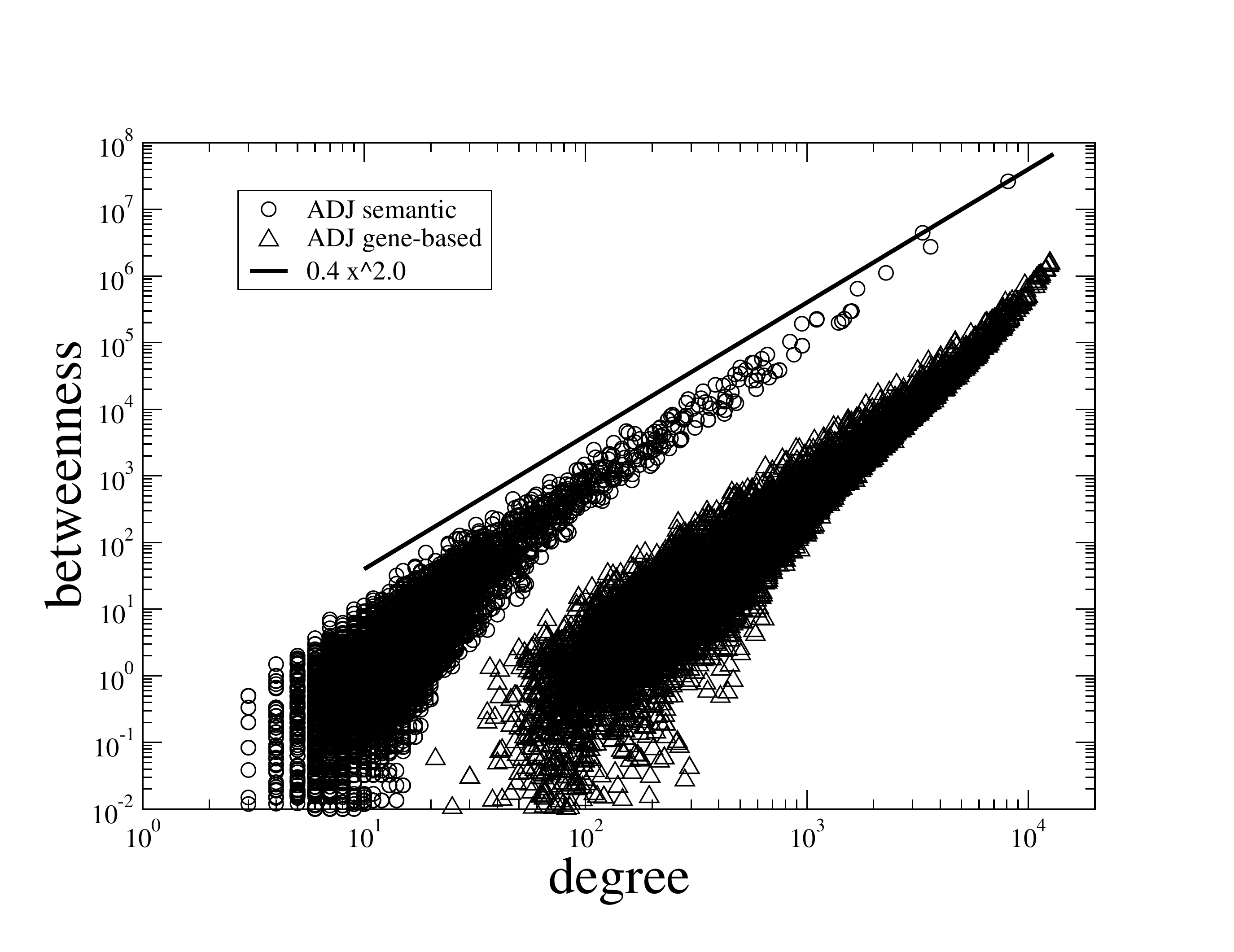} 
              }
              \caption{The left panel shows a comparison between the degree distribution for the semantic (circles) and gene-based (triangles) networks. The right panel shows the scatter-plot describing the relationships between degree and betweenness  for each term in the adjacency semantic (circles) and gene-based (triangles) networks. The solid line represent a reference curve $y \propto x^2$ showing the both scatter-plots are compatible with a quadratic dependance of the betweenness from the degree.} \label{GOnetdegree}
\end{center}
\end{figure}

In Fig. \ref{GOnetdegree} (left panel) we show the degree distribution \cite{R} for the semantic (circles) and gene-based (triangles) networks. The profile of the two distributions is quite different, mainly due to the fact that terms in the gene-based network are much more linked to each other. This might be an indication of the fact that genes perform different tasks within different biological processes, molecular functions and cellular components. Moreover, it should also be noted that this might be an artifact of the fact that when generating the gene-based network a gene annotated in term T is also assigned to all terms that are parents of T. In Fig. \ref{GOnetdegree} (right panel) we show the scatter-plot describing the relationships between degree and betweenness  \cite{R} for each term in the semantic (circles) and gene-based (triangles) networks. Contrary to the degree distribution, this panel shows that the two scatter-plots are quite similar. The solid line represents a reference curve $y \propto x^2$ showing the both scatter-plots are compatible with a quadratic dependance of the betweenness from the degree. These results suggest that the main difference between the two networks is in the number of links and not in the relative importance of the terms within the network. In this respect, it is worth mentioning that the average path length \cite{R} of the semantic adjacency network is 1.997 while the average path length of the adjacency gene-based network is 1.935 that is very close.

The gene-based network is fully connected \cite{partitioning}. Therefore even terms belonging to different branches of the GO (namely {\sl{GO:0008150 (Biological Processes)}}, {\sl{GO:0005575 (Cellular Component)}} and {\sl{GO:0003674 (Molecular Function)}}) can be linked together when looking at their gene-content. The number of links between the 8118 terms of the BP branch is 3054021. The number of links between the 3336 terms of the MF branch is 106532 and the number of links between the 1110 terms of the MF branch is 62339, as summarized in Table \ref{tabcompareADJ}. The number of links connecting terms of the different branches is 1919851, i.e.  37\% of the total number of links. The majority of links therefore connects terms of the BP branch although a relevant number of links also connects the BP branch with the others. The number of links in the gene-based network is larger than in the  semantic network, and this is maintained when one restricts the analysis to the three main branches, see Table \ref{tabcompareADJ}. This probably explains why the average path length is on average slightly smaller in the gene-based network. It is perhaps surprising the fact that the difference in the average  path lengths is so small, despite the fact that the difference in the number of links is so large between the two networks. This might be an indication of redundancy: the relationships between biological processes mediated by genes might go through many different channels in order to ensure that a link between the terms is always active, despite possible impairments of some of the channels.  
\begin{table}
\begin{tabular}{|c||c|c|c||c|c|c|}
\hline
network~ & nodes & links  & APL& nodes & links & APL \\
 \hline
  ALL & 12564 &116422 & 1.997& 12564& 5142743 & 1.935 \\ 
    BP &   8118 &   85447 & 1.997 &   8118 & 3054021& 1.907\\
    MF &   3336 &  18489  & 1.997 &   3336 &   106532 &1.981\\ 
    CC &   1110 &  12486  & 1.980 &   1110 &     62339 & 1.899\\  \hline
\end{tabular}  
\caption{Basic metrics for the adjacency gene-based and semantic network. In the table we also show the results for the three main in GO branches,  namely {\sl{GO:0008150 (Biological Processes)}}, {\sl{GO:0005575 (Cellular Component)}} and {\sl{GO:0003674 (Molecular Function)}. These branches are disconnected in the semantic network, while they are linked in the gene-based network. APL stands for Average path Length.}} \label{tabcompareADJ}
\end{table}

\subsection{The Bonferroni gene-based network} \label{bipartite}

As we mentioned above, the adjacency gene-based network is fully connected and shows an high level of redundancy. One might therefore ask what are the main links between terms. The answer can be given by considering statistically validated networks. Such networks only involve links that are statistically validated against a null hypothesis of randomness that takes into account the natural heterogeneity of the system. Therefore they can be considered as a tool for filtering relevant information out of the system, based on the terms gene content.

Let us consider here the Bonferroni network of GO terms. Given the large heterogeneity in the genes set, we have constructed the validated network by adopting the validation procedure that involves the construction of subsystems where all elements have the same degree. The Bonferroni network of GO terms thus obtained involves 558 nodes and 3508 links. It then involves the 4.75\% of terms and the 0.08\% of links with respect to the adjacency network. These numbers testify how large the reduction of both nodes and links can be when the statistical significance of a link is assessed by using a null hypothesis that properly takes into account the heterogeneity of the system \cite{FFDDRR}. A suggestive explanation of such large reduction is again redundancy: the fact that the adjacency gene-based network contains so many links compatible with a null hypothesis of randomness might be due to the need of creating as many channels  of communication amongst terms as possible so that impairments have negligible impact on the system. In this respect, the Bonferroni network provides the core of the system.

\begin{figure} 
\begin{center}
             {
              \includegraphics[scale=0.30] {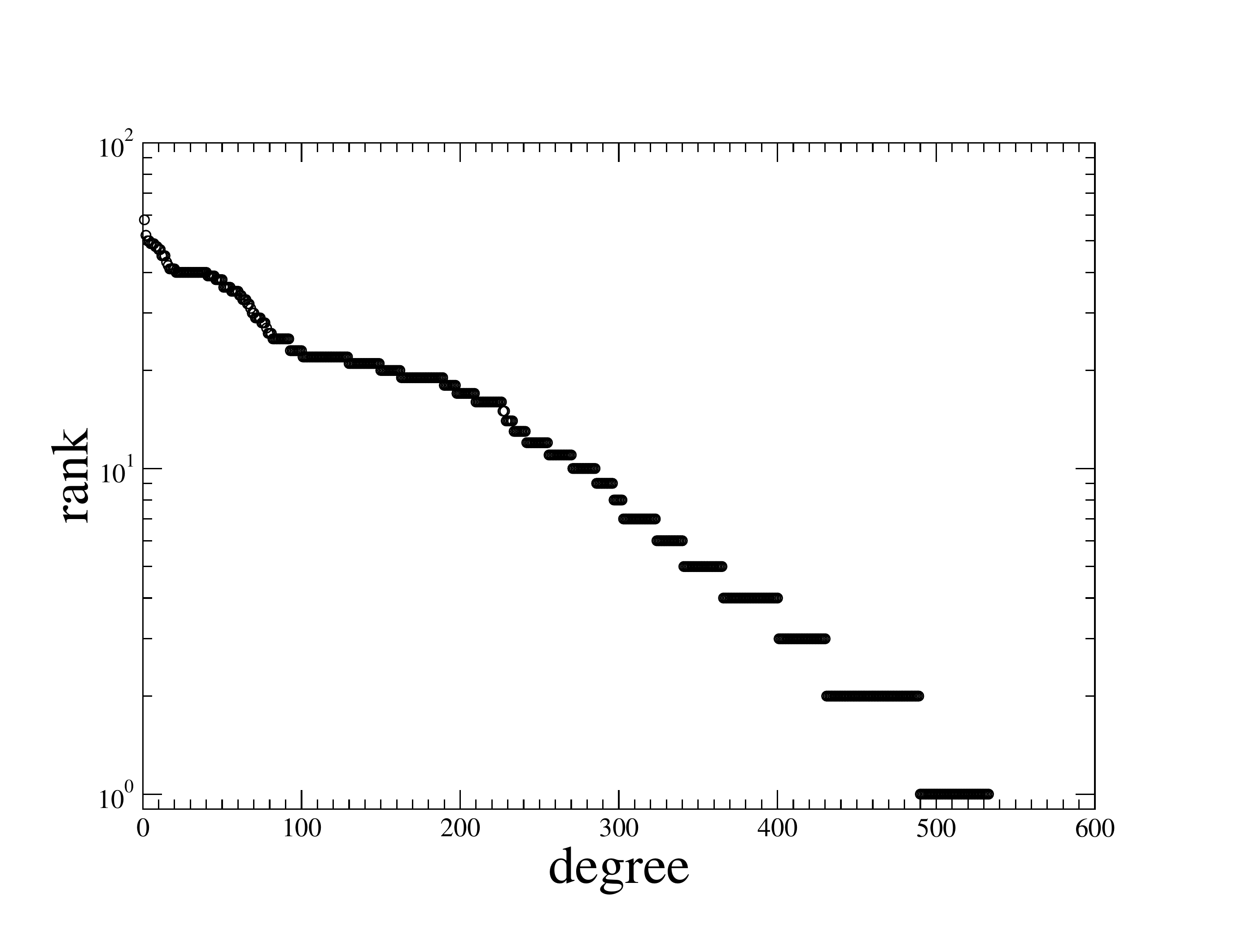} 
              \includegraphics[scale=0.30] {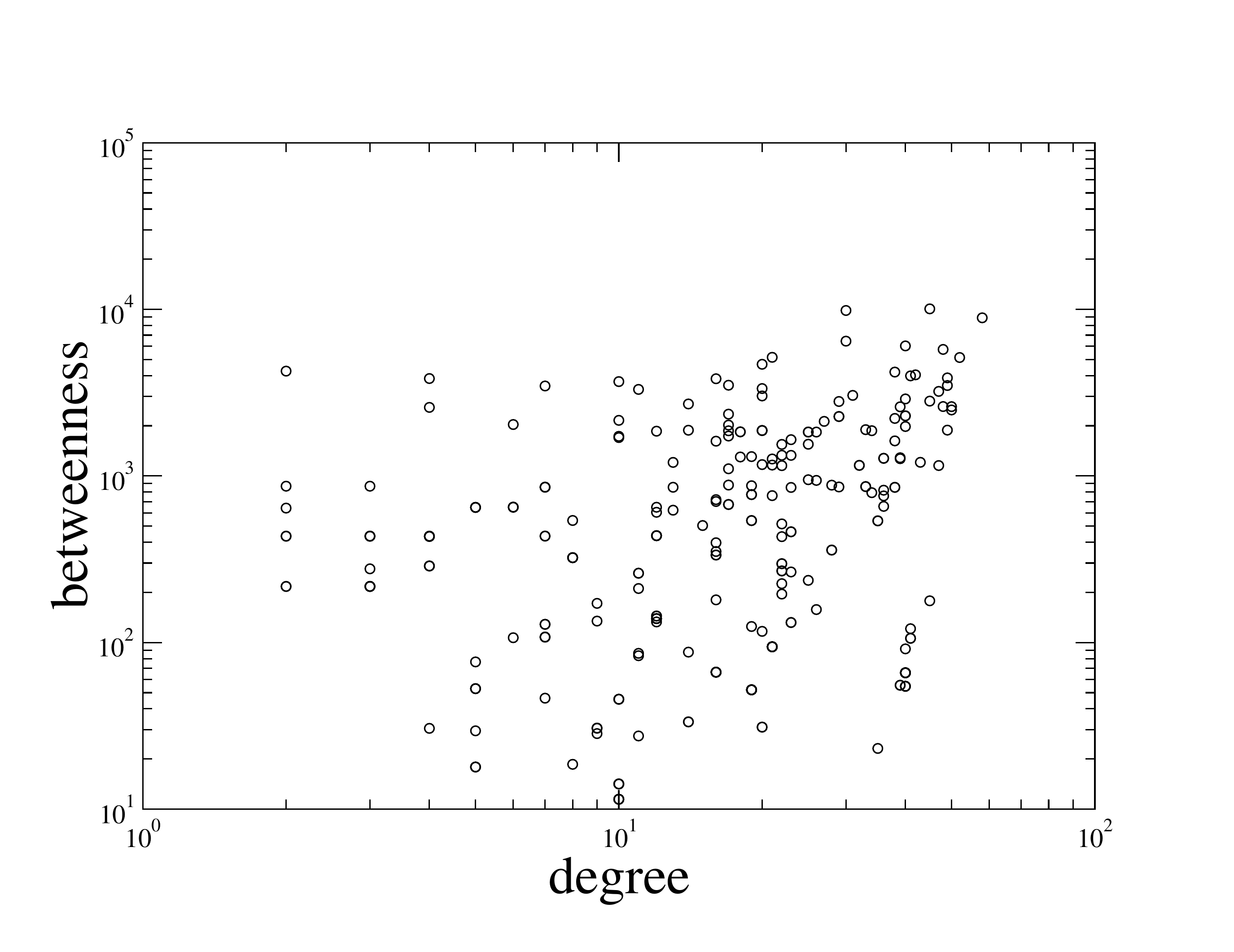} 
             } 
              \caption{The left panel shows the degree rank plot for the Bonferroni gene-based network. The right panel shows the scatter-plot describing the relationships between degree and betweenness  for each term in the the Bonferroni gene-based network.} \label{BONFdeg}
\end{center}
\end{figure}

In Fig. $(\ref{BONFdeg})$ (left panel) we show the degree \cite{R} of the nodes present in the Bonferroni network. The most connected nodes have degree values over 50. This means that despite the large reduction of nodes with respect to the adjacency network, there are still nodes that behave like hubs. In Fig. $(\ref{BONFdeg})$ (right panel) we show the scatter-plot describing the relationships between degree and betweenness  \cite{R} for each term in the Bonferroni network. When comparing this figure with Fig. $(\ref{GOnetdegree})$ (panel b) the severe cut in the number of nodes and links becomes evident. This is confirmed when computing the average path length \cite{R} for the Bonferroni network that amounts to 4.017.

\begin{figure} 
\begin{center}
              {
                \includegraphics[scale=0.24] {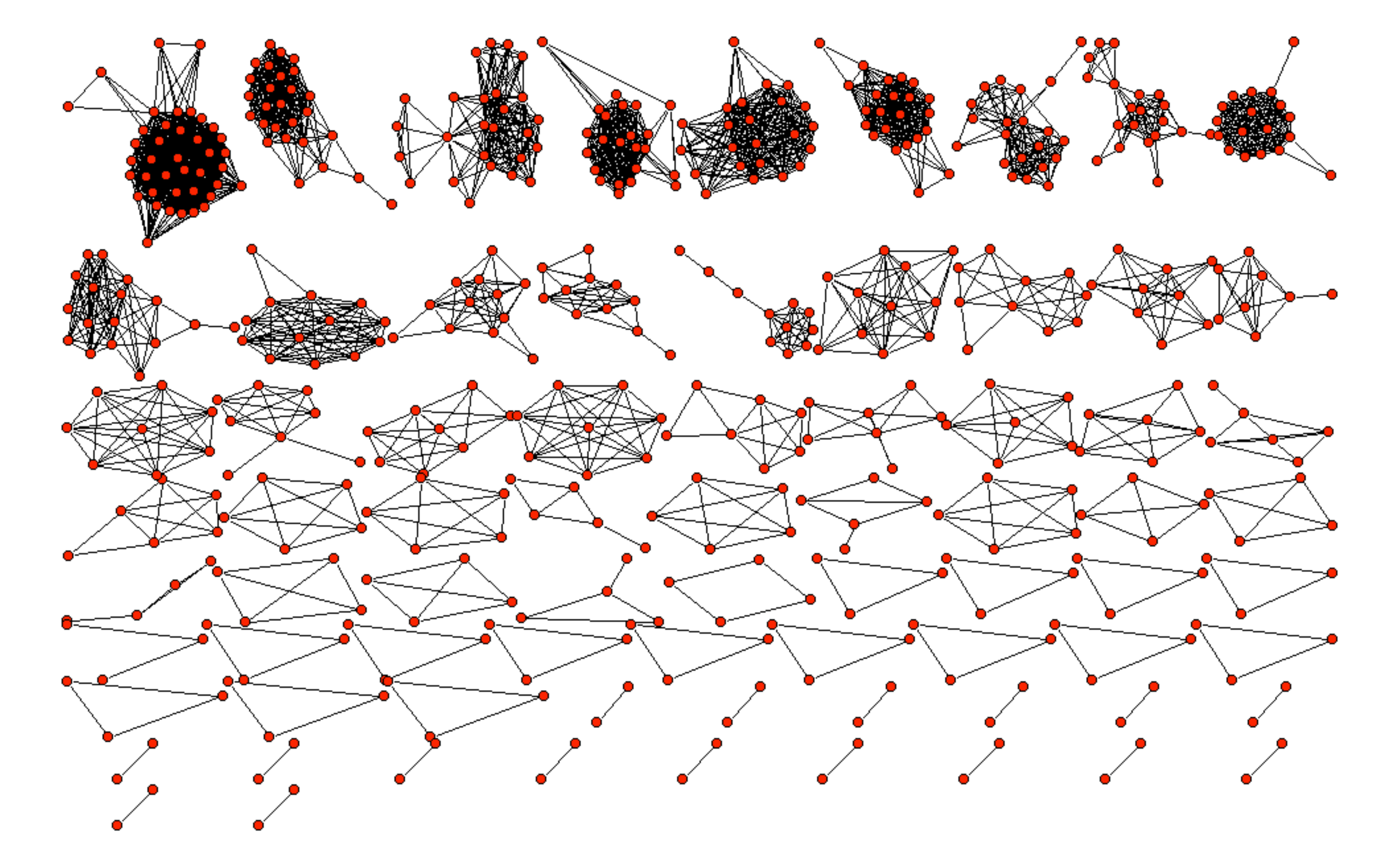} 
                \includegraphics[scale=0.26] {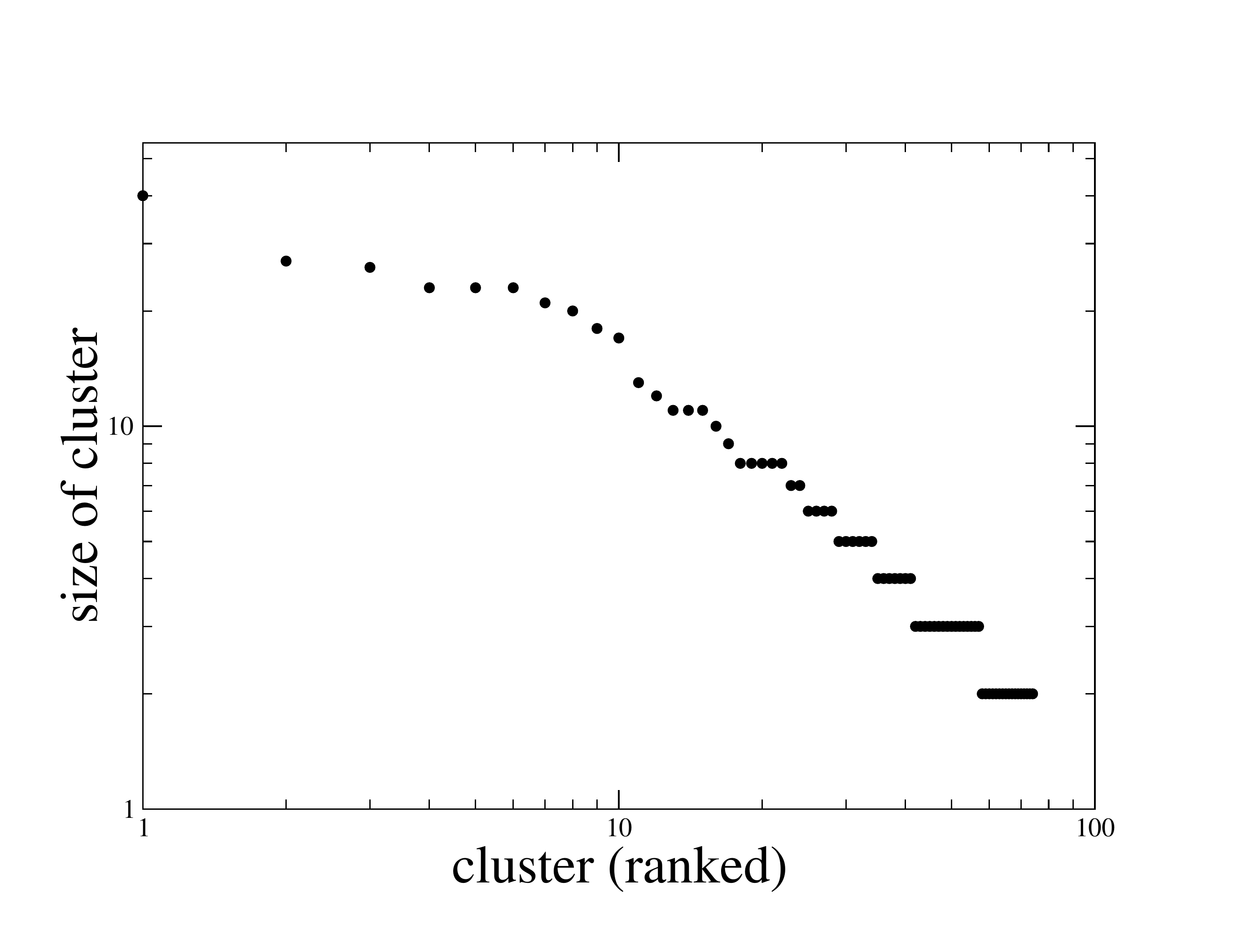} 
              }
              \caption{Partition of the gene-based bonferroni network by using the Infomap algorithm \cite{Rosvall2008}. The left panel shows the different communities observed. The right panel shows a rank plot of the size of communities.} \label{BONFsizeclus}
\end{center}
\end{figure}

The Bonferroni network of GO terms is naturally partitioned in 30 clusters. The largest cluster has size 467. The second largest has size 8, thus indicating the presence of a giant component. We also partitioned the network by using the Infomap algorithm \cite{Rosvall2008}. As a result, we get a partition involving $N_p^{(g)}=74$ clusters whose size ranges from 2 to 40, as shown in Fig. \ref{BONFsizeclus}.

\subsection{Community characterization of the Bonferroni gene-based network} \label{netcomm2}

We can therefore compare the information encoded in the semantic GO structure with the one associated to the genes annotated in the terms at the level of network communities. In fact, each of the above clusters groups together GO terms on the bases of their gene content. However, any GO terms T in a gene-based cluster C carries its semantic information inherited from the general GO semantic structure. We want to bring together these two levels of information by using the semantic information of any GO term T to characterize the gene-based cluster C that includes T. The idea is to use the methodology of Ref. \cite{cluschar} and to use the semantic information of T as attribute for the characterization of gene-based cluster C. Specifically, we will use the semantic  information obtained in the Infomap partitioning of the adjacency semantic network. We remember that in section \ref{comparison} we obtained a partitioning of the adjacency semantic network in communities that are homogeneous from a biological point of view. Thus, we here considered as attributes the clusters $A_i, i=1, \cdots, 163$ of Fig. \ref{ADJsemnet}. Each cluster $C$ of GO terms in the gene-based Bonferroni network can be therefore characterized in terms of $163$ attributes. In Table  \ref{tab2} we report the over-expressions for the first largest clusters, with size greater than 20. The full list of characterizations is given in Table SM3 of the Supplementary Material. 

\begin{table}
\begin{tabular}{|c|c||l|}
\hline
Cluster~ & size & overexpressed attribute\\
 \hline
$  4$ & $40$&{\sl{Cluster 8 (BP terms)}}\\ \hline
$  7$ & $27$&{\sl{Cluster 8 (BP terms)}}\\
$  7$ & $27$&{\sl{Cluster 12 (CC terms)}}\\ \hline
$17$ & $26$&{\sl{Cluster 42 (BP terms)}}\\ 
$17$ & $26$&{\sl{Cluster 39 (CC terms)}}\\ \hline
$15$ & $23$&{\sl{Cluster 44 (BP terms)}}\\ 
$15$ & $23$&{\sl{Cluster 17 (BP terms)}}\\
$15$ & $23$&{\sl{Cluster 14 (BP terms)}}\\ \hline
$13$ & $23$&{\sl{Cluster 60 (BP terms)}}\\ \hline
$  3$ & $23$&{\sl{Cluster 38 (BP terms)}}\\  
$  3$ & $23$&{\sl{Cluster 23 (CC terms)}}\\  \hline
$11$ & $21$&{\sl{Cluster 51 (BP terms)}}\\ 
$11$ & $21$&{\sl{Cluster 66 (CC terms)}}\\ \hline
\end{tabular} 
\caption{Statistical characterization of the communities of the Bonferroni gene-based network. Communities have been obtained by using the Infomap algorithm \cite{Rosvall2008}. The characterization has been done by using the methodology of Ref. \cite{cluschar}. For each term T we have considered as attribute the membership of T to one of the communities of the adjacency semantic network. Only the results for the five largest clusters are shown. The full list of characterizations is given in Table SM3 of the Supplementary Material.} \label{tab2}
\end{table}

One first thing to be noticed is that clusters can be characterized in terms of more than one attribute. Moreover, Table \ref{tab2} shows that in the same clusters of the gene-based Bonferroni network one can have terms that belong to two different  GO branches. These examples show how there exist GO terms that have no semantic link between each other and can nevertheless be put in connection when the gene content of their children terms is considered. In other words, if we think to the attributes as defining homogeneous and distinct subsets of the GO, such subsets might be disconnected from a semantic point of view and connected when the gene content of their terms is taken into account.

\section{Conclusion} \label{concl}

In this paper we have considered the Gene Ontology from a complex network perspective. We have considered two types of GO networks: the one associated to the semantic structure of the Gene ontology and the one associated to the gene content of each GO term.

We have first considered some basic network metrics showing that  the main differences between the two networks are in the number of links and not in the relative importance of the terms within the network.

We have then compared the two networks at the level of network communities. In the adjacency semantic network we detected 163 communities by using the Infomap algorithm. Such communities are homogeneous from a biological point of view. This has been confirmed by a statistical analysis able to provide the 3-words that are over-expressed in the obtained clusters. In the Bonferroni gene-based network we detected 74 small communities by using the Infomap algorithm. Our results show that there exist GO terms that have no semantic link between each other and can nevertheless be put in connection when the gene content of their children terms is considered.

We believe that the importance of our results is twofold. On one side we have devised a simple methodology, based on the detection of the statistical significant 3-words, that allows to have a first glance insight about the biological meaning of groups of GO terms. This might become a routinary tool able to provide a first-glance biological interpretation of groups of GO terms. On the other side we have shown that a deeper analysis of GO terms, based on their gene content, might reveal relationships between terms that are missed by looking at the semantic structure of GO. This has a practical importance for the evolution and maintenance of GO. In fact this kind of analysis can be useful to capture the relations amongst genes to be profitably transferred at the level of terms. However, this is also important from a biomedical point of view, as it might reveal how genes over-expressed in a certain term also affect other biological processes, molecular functions and cellular components not directly linked by the GO semantics.




\begin{thebibliography}{}
\bibitem{FFDDRR}
A less restrictive filtering is done when considering the FDR network. In this case one would have 915 nodes and 7670 links.
\end{thebibliography}
\end{document}